\documentclass[usegraphicx,useAMS,usenatbib,]{mn2e}
\usepackage{aas_macros}

\newcommand{\Teff}{\mbox{${\rm T}_{\rm eff}$}}
\newcommand{\Tspec}{\mbox{${\rm T}_{\rm spec}$}}
\newcommand{\Tirfm}{\mbox{${\rm T}_{\rm IRFM}$}}
\newcommand{\logg}{\mbox{$\log\,{\rm g}$}}
\newcommand{\ubvric}{\mbox{UBV(RI)$_{\rm C}$}}
\newcommand{\Fearth}{\mbox{${\cal F}_{\oplus}$}}
\newcommand{\halpha}{\mbox{{\rm H}$_{\alpha}$}}

\title[UBV(RI)$_{\rm C}$ photometry of transiting planet host stars]
  {UBV(RI)$_{\rm C}$ photometry of transiting planet host stars}
\author[P.F.L. Maxted et~al.]
{P.F.L.~Maxted$^1$, Koen, C.$^2$, Smalley, B.$^1$\\
  $^1$Astrophysics Group,  Keele University, Keele, 
      Staffordshire ST5 5BG, United Kingdom\\
 $^2$Department of Statistics, University of the Western Cape, Private Bag
X17, Bellville 7535, South Africa
}
\date{Submitted 2011}

\pagerange{\pageref{firstpage}--\pageref{lastpage}} \pubyear{2008}

\def\LaTeX{L\kern-.36em\raise.3ex\hbox{a}\kern-.15em
    T\kern-.1667em\lower.7ex\hbox{E}\kern-.125emX}

\begin{document}
\label{firstpage}

\maketitle

\begin{abstract}
 We present new \ubvric\ photometry of 22 stars that host transiting planets,
19 of which were discovered by the WASP survey. We use these data together
with 2MASS JHK$_{\rm S}$ photometry to estimate the effective temperature
of these stars using the infrared flux method. We find that the effective
temperature estimates for stars discovered by the WASP survey based on the
analysis of spectra are reliable to better than their quoted uncertainties. 
\end{abstract}


\section{Introduction}

 The transit of a planet across the face of its host star provides us with the
opportunity to measure the properties of the planet in great detail. Essential
to exploiting this opportunity is a good understanding of the host star
itself. For example, a combined analysis of the transit lightcurve together
with the spectroscopic orbit of the host star leads directly to a measurement
of the host star density, $\rho_{\star}$, and the surface gravity of the
planet \citep{ 2007MNRAS.379L..11S,  2003ApJ...585.1038S}. To estimate the
mass and radius of the planet, an additional constraint is needed. The details
of how the additional constraint is applied varies between different groups,
but they generally share the common feature that an analysis of the
spectrum is used to estimate the stellar effective temperature, \Teff, stellar
surface gravity, \logg, and metallicity, [Fe/H]\footnote{[Fe/H] is the iron
abundance relative to the Sun, metallicity is normally estimated by assuming
that the abundances of other elements scales with the iron abundance}, and
these are combined with the estimate of $\rho_{\star}$ to estimate the mass
and radius of the star using either an empirical calibration or stellar models
\citep{2010MNRAS.408.1689S,  2010A+A...516A..33E,2010ApJ...710.1724B}. The
radius and mass of the planet then follow directly from the observed 
depth of the transit and Kepler's Law, respectively.

 Irrespective of the details, it is clear that the accurate characterisation
of the host star is essential for an accurate understanding of the planets
that orbit it. Most transiting planets discovered to-date have been found
using wide-angle, ground-based photometric surveys such as WASP
\citep{2006PASP..118.1407P} and  HATNet \citep{2004PASP..116..266B}. These
surveys target stars with visual magnitudes in the approximate range
8.5\,--\,13.  One obstacle to the  accurate characterisation of these stars
is the poor quality of the optical photometry that is generally available
for stars of this brightness. 

 Accurate photometry for bright stars is available at infrared wavelengths
(JHK$_{\rm S}$) across the entire sky from the 2MASS survey
\citep{2006AJ....131.1163S} and in the southern hemisphere from the 
DENIS  survey \citep{2005yCat.2263....0T}. The DENIS survey extends the
available photometry to the I-band. In the northern hemisphere the CMC14
catalogue provides r'-band photometry \citep{2002A&A...395..347E}. In the
optical regime, BV photometry is available for stars brighter than
$V\approx12$ from the Tycho-2 catalogue \citep{2000A&A...355L..27H}, although
this catalogue is only complete to $V\approx11$ and the photometric precision
deteriorates rapidly for  $V\ga9.5$. Accurate optical photometry provides flux
measurements around the peak of the spectral energy distribution (SED) for
solar-type stars, unlike infrared photometry that samples the Rayleigh-Jeans
tail of the SED. Optical photometry is also sensitive to reddening and
metalicity, both of which affect the blue end of the SED much more than the
red end.  Both these effects need to be accurately accounted for if the
distance to the star is to be estimated from the photometric properties of the
star, e.g., for kinematical studies. The combination of accurate optical and
infrared photometry also makes it possible to make a robust and accurate
estimate of the star's effective temperature using the infrared flux method.
(\citealt{1979MNRAS.188..847B}). Accurate optical photometry is also useful
for planning follow-up observations, e.g., for estimating optimum exposure
times.

 In this paper we present new, high quality photoelectric optical photometry
for 22 planet-host stars, mostly WASP discoveries in the southern hemisphere.
We use this photometry to make an independent check on the accuracy of the
effective temperature estimates published for these stars based on the
analysis of their spectra.

\section{Observations}

 Observations were obtained with the SAAO 0.5-m telescope and modular
photometer \citep{1988MNSSA..47...69K}. This is a very stable and well
understood instrumental setup for obtaining standardised \ubvric\ photometry 
\citep{2005ARA+A..43..293B}. Observations were obtained in dark sky conditions
over the course of two observing runs, 2010 September 9-12 and 2011 March
5-12. Reduction of the instrumental magnitudes to the standard photometric
system defined by the E-region standards of \cite{1989SAAOC..13....1M}
followed the methods described in the appendix of \cite{1988MNSSA..47...69K}.
The observations are presented in Table~\ref{phottable}.

\begin{table}
\caption{Photometry of 22 transiting planet host stars. $N$ is the number of
observations obtained. Standard errors are given for stars with 3 or more
observations.}
\label{phottable}
\begin{tabular}{@{}lrrrrrr}
\hline
Star 
& \multicolumn{1}{c}{V} 
& \multicolumn{1}{c}{B$-$V}
& \multicolumn{1}{c}{U$-$B} 
& \multicolumn{1}{c}{V$-$R }
& \multicolumn{1}{c}{V$-$I }
& $N$\\
\hline
WASP-2  & 11.788 & 0.897 & 0.604 & 0.488 & 0.932 & 2 \\
WASP-4  & 12.463 & 0.783 & 0.362 & 0.405 & 0.782 & 2 \\
WASP-5  & 12.136 & 0.705 & 0.281 & 0.380 & 0.727 & 2 \\
WASP-7  & 9.483 & 0.460 & 0.015 & 0.267 & 0.528 & 2 \\
WASP-8  & 9.773 & 0.747 & 0.369 & 0.404 & 0.788 & 2 \\
\noalign{\smallskip}
WASP-15         &  10.918 & 0.495 &$ 0.008 $& 0.295 & 0.579 & 3 \\
                &   $\pm$0.003 & 0.001 &  0.008  & 0.013 & 0.012 &   \\
\noalign{\smallskip}
WASP-16 & 11.309 & 0.741 & 0.283 & 0.380 & 0.763 & 2 \\
WASP-17 & 11.500 & 0.496 & 0.040 & 0.268 & 0.585 & 2 \\
WASP-18 & 9.273 & 0.484 & 0.013 & 0.278 & 0.548 & 2 \\
\noalign{\smallskip}
WASP-19         &     12.312 & 0.785 &$ 0.398$& 0.425 & 0.812 & 4  \\
                & $\pm$0.017 & 0.004 &  0.052 & 0.007 & 0.009 &    \\
\noalign{\smallskip}
WASP-22 & 11.708 & 0.603 & 0.139 & 0.323 & 0.638 & 2 \\
WASP-25 & 11.848 & 0.727 & 0.253 & 0.380 & 0.764 & 2 \\
WASP-26 & 11.099 & 0.621 & 0.145 & 0.344 & 0.690 & 2 \\
WASP-28 & 12.148 & 0.596 & 0.033 & 0.329 & 0.690 & 2 \\
WASP-29 & 11.207 & 1.087 & 1.061 & 0.620 & 1.119 & 2 \\
\noalign{\smallskip}
WASP-31         &     11.937 & 0.513 &$-0.009$& 0.297 & 0.593 & 3 \\
                & $\pm$0.006 & 0.017 &   0.004& 0.016 & 0.004 &   \\
\noalign{\smallskip}
WASP-34         &     10.366 & 0.684 &  0.224 & 0.364 & 0.716 & 3 \\
                & $\pm$0.012 & 0.003 &  0.011 & 0.013 & 0.015 &   \\
\noalign{\smallskip}
WASP-37$^{\rm a}$ & 12.717 & 0.628 & 0.022 & 0.337 & 0.699 & 2 \\
\noalign{\smallskip}
WASP-39         &  12.100 & 0.803 &  0.370  & 0.426 & 0.852 & 3 \\
                &   $\pm$0.012 & 0.021 & 0.040   & 0.006 & 0.021 &   \\
\noalign{\smallskip}
HAT-P-24 & 11.754 & 0.462 & $-0.017$ & 0.260 & 0.518 & 2 \\
\noalign{\smallskip}
HAT-P-27/       & 12.163 & 0.909 & 0.645 & 0.467 & 0.892 & 3 \\
WASP-40        &  $\pm$0.011 & 0.010 & 0.009 & 0.009 & 0.016 &   \\
\noalign{\smallskip}
CoRoT-7$^{\rm b}$ &     11.718 & 0.849 &$ 0.915$ & 0.437 & 0.827 & 3 \\
                &   $\pm$0.003 & 0.026 &         & 0.012 & 0.027 &   \\
\hline
\end{tabular}
\medskip
$^{\rm a}$One discrepant measurement ignored
$^{\rm b}$Only one reliable U-band measurement was obtained. \newline
\end{table}

\section{Analysis}

 We estimate the effective temperature of these stars using a simplified
version of the infrared flux method (IRFM; \citealt{1979MNRAS.188..847B}).
The essence of this method is to find values of the effective temperature,
\Teff, and angular diameter, $\theta$, for which stellar atmosphere models
simultaneously satisfy the observed value of the bolometric flux at
the Earth, \Fearth, and the observed flux at infrared wavelengths. Photometry
covering the peak of the SED is important for an accurate estimate of \Fearth\
in solar-type stars. Some type of interpolation  scheme is required to enable
the integration of an SED that is only  sparsely sampled by broadband
photometry. In our method we use numerical integration of the best-fitting
model SED from a grid of  stellar atmosphere models. The infrared flux of a
solar-type star predicted by stellar models is insensitive to parameters such
as surface gravity, metalicity and  reddening or to the details of the model,
so the value of \Teff\ is almost model independent. In principle, the value of 
\Teff\ derived by our method could be improved by recalculating or
interpolating the model SED used in the integration of \Fearth\ and iterating
this process. In practice we find that further refinement of our \Teff\
estimate is not required. An estimate of the interstellar reddening is
required for an accurate comparison of the model SED to the observed fluxes.
In practice, we find that the effect of interstellar reddening is negligible
for the majority of the stars we have studied.

 We use the data provided by the 2MASS survey \citep{2006AJ....131.1163S}   to
obtain the JHK$_{\rm S}$ magnitudes for our targets. For stars where we have
access to the spectra, we have measured the equivalent width of the
interstellar absorption features due to sodium near 589\,nm and used the
calibration of \cite{1997A+A...318..269M} to convert this to an estimate of
the interstellar reddening, E(B$-$V). We estimate the bolometric flux
from the star by integrating the best-fitting model spectral energy
distribution (SED) from grid of models at 250K intervals in \Teff\ from
\cite{1993KurCD..13.....K}. We use least-squares fitting of the model fluxes
integrated over the appropriate bandpasses for all available optical and
infrared magnitudes to determine the best-fitting model SED.  For stars with 3
or more observations we use the standard errors on the means to determine the
weights of the \ubvric\ data in the fit. We use linear regression on the
apparent magnitudes and standard errors in each bandpass for these stars to
determine a relationship between apparent magnitude and standard error for
each bandpass, and then use this to assign a weight to the data for stars with
fewer than 3 observations. For the conversion of magnitudes to fluxes we use
the zero-point values from \citet{1979PASP...91..589B} and
\citet{2003AJ....126.1090C}.

  We include the effect of interstellar reddening in the least-squares fit of
the grid of model SEDs for stars where an estimate of E(B$-$V) from the
spectrum is available. The spectral energy distributions were de-reddened
using the analytical extinction expressions from \citet{1983MNRAS.203..301H}.
For the two stars for which we do not have access to the spectra we assume
E(B$-$V)$=0$.  For a typical star with $\Teff \approx6000$\,K and
E(B$-$V)$\approx$0.01, neglecting the reddening results in an over-estimate of
\Teff\ by about 30\,K.  The effective temperature of the star is then
estimated using each of the available J, H and K$_{\rm S}$ magnitudes
independently and the weighted average of these results is taken as the final
value. We refer to this estimate of the effective temperature as \Tirfm.

 The values of E(B$-$V) and \Tirfm\ are given in Table~\ref{tefftable}
together with published estimates of the effective temperature for each star
based on the analysis of the spectrum, \Tspec.

\begin{table*}
\caption{Effective temperature estimates for our target stars. The reddening
E(B-V) is estimated from the equivalent width of the interstellar sodium `D'
absorption lines. }
\label{tefftable}
\begin{tabular}{@{}lrrrrll}
\hline
Star& E(B$-$V)  & \multicolumn{1}{l}{T$_{\rm IRFM}$/K} &
\multicolumn{1}{l}{T$_{\rm Spec}$/K}  &
\multicolumn{1}{l}{[Fe/H]} & Spectrograph & Reference \\
\hline
WASP-2    &  0.02 & 5110 $\pm$  60 & 5200 $\pm$ 200 & $\sim 0.0$    & SOPHIE  & \cite{2007MNRAS.375..951C} \\
          &       &                & 5150 $\pm$ ~80 &$-0.08$ $\pm$  0.08& HARPS   & \cite{2010A+A...524A..25T} \\
WASP-4    &  0.00 & 5540 $\pm$  55 & 5500 $\pm$ 150 &$ 0.0 $ $\pm$  0.2 & CORALIE & \cite{2008ApJ...675L.113W}\\
          &       &                & 5500 $\pm$ 100 &$-0.03$ $\pm$ 0.09 & UVES    & \cite{2009A+A...496..259G} \\
WASP-5    &  0.00 & 5770 $\pm$  65 & 5700 $\pm$ 100 &$ 0.0 $ $\pm$ 0.2  & CORALIE & \cite{2008MNRAS.387L...4A} \\
          &       &                & 5700 $\pm$ 100 &$ 0.09$ $\pm$ 0.09 & UVES    & \cite{2009A+A...496..259G} \\
WASP-7    &  0.00 & 6520 $\pm$  70 & 6400 $\pm$ 100 &$ 0.0 $ $\pm$ 0.1  & CORALIE & \cite{2009ApJ...690L..89H} \\
WASP-8    &  0.00 & 5570 $\pm$  85 & 5600 $\pm$ ~80 &$0.17 $ $\pm$ 0.17 & HARPS   & \cite{2010A+A...517L...1Q} \\
WASP-15   &  0.00 & 6210 $\pm$  60 & 6300 $\pm$ 100 &$1.4  $ $\pm$ 0.1  & CORALIE & \cite{2009AJ....137.4834W} \\
WASP-16   &  0.01 & 5550 $\pm$  60 & 5700 $\pm$ 150 &$0.01 $ $\pm$ 0.10 & CORALIE & \cite{2009ApJ...703..752L} \\
WASP-17   &  0.05 & 6500 $\pm$  75 & 6550 $\pm$ 100 &$-0.25$ $\pm$ 0.09 & CORALIE & \cite{2010ApJ...709..159A} \\
          &       &                & 6650 $\pm$ ~80 &$-0.19$ $\pm$ 0.09 & HARPS   & \cite{2010A+A...524A..25T} \\
WASP-18   &  0.00 & 6455 $\pm$  70 & 6400 $\pm$ 100 &$0.00 $ $\pm$ 0.09 & HARPS   & \cite{2009Natur.460.1098H} \\
WASP-19   &  0.00 & 5440 $\pm$  60 & 5500 $\pm$ 100 &$0.02 $ $\pm$ 0.09 & CORALIE & \cite{2010ApJ...708..224H} \\
WASP-22   &  0.01 & 6020 $\pm$  50 & 6000 $\pm$ 100 &$0.05 $ $\pm$ 0.08 & HARPS   & \cite{2010AJ....140.2007M} \\
WASP-25   &  0.00 & 5615 $\pm$  55 & 5750 $\pm$ 100 &$-0.05$ $\pm$ 0.10 & CORALIE & \cite{2011MNRAS.410.1631E} \\
WASP-26   &  0.01 & 6015 $\pm$  55 & 5950 $\pm$ 100 &$-0.02$ $\pm$ 0.09 & CORALIE & \cite{2010A+A...520A..56S} \\
WASP-28   &  0.04 & 6190 $\pm$  60 & 6100 $\pm$ 150 &$-0.29$ $\pm$ 0.10 & CORALIE & \cite{West2010} \\
WASP-29   &  0.00 & 4875 $\pm$  65 & 4800 $\pm$ 150 &$ 0.11$ $\pm$ 0.14 & CORALIE & \cite{2010ApJ...723L..60H} \\
WASP-31   &  0.00 & 6175 $\pm$  70 & 6250 $\pm$ 150 &$-0.29$ $\pm$ 0.11 & CORALIE & \cite{2010arXiv1011.5882A} \\
          &       &                & 6300 $\pm$ 100 &$-0.20$ $\pm$ 0.09 & HARPS   & \cite{2010arXiv1011.5882A} \\
WASP-34   &  0.00 & 5695 $\pm$  65 & 5700 $\pm$ 100 &$-0.02$ $\pm$ 0.10 & CORALIE & \cite{2011A+A...526A.130S} \\
WASP-37   &  0.05 & 5940 $\pm$  55 & 5800 $\pm$ 150 &$-0.40$ $\pm$ 0.12 & CORALIE+SOPHIE & \cite{2011AJ....141....8S} \\
WASP-39   &  0.04 & 5460 $\pm$  55 & 5400 $\pm$ 150 &$-0.12$ $\pm$ 0.10 & CORALIE & \cite{2011arXiv1102.1375F} \\
HAT-P-24  &       & 6330 $\pm$  65 & 6373 $\pm$ ~80 &$-0.16$ $\pm$ 0.08 & HIRES   & \cite{2010ApJ...725.2017K} \\
HAT-P-27  &       & 5175 $\pm$  70 & 5300 $\pm$ ~90 &$0.29 $ $\pm$ 0.10 & HIRES   & \cite{2011arXiv1101.3511B} \\
=WASP-40  &  0.01 &                & 5200 $\pm$ 150 &$0.14 $ $\pm$ 0.11 & CORALIE+SOPHIE & \cite{2011arXiv1101.4643A} \\
CoRoT-7   &       & 5240 $\pm$  55 & 5275 $\pm$ ~75 &$0.03 $ $\pm$ 0.06 & UVES & \cite{2009A+A...506..287L} \\
          &       &                & 5250 $\pm$ ~60 &$0.12 $ $\pm$ 0.06 & HARPS+UVES & \cite{2010A+A...519A..51B} \\
\hline
\end{tabular}
\end{table*}

\section{Discussion}
 For all 25  stars from the WASP survey a method based on an analysis of the
\halpha\ line has been used to measure \Tspec\ \citep{2011A+A...526A.130S}.
There may be systematic errors in these \Tspec\ estimates due to instrumental
effects such as scattered light, the normalization of the spectra, etc., as
well as systematic errors in the model stellar atmospheres used to analyse the
\halpha\ line. Similar issues will affect the \Tspec\ estimates for stars from
other sources. \citet{2010MNRAS.405.1907B} have compared \Tspec\ estimates for
23 nearby solar-type stars to \Teff\ determined directly from interferometric
angular diameters. They find that their \Tspec\ estimates are too hot by
$40\pm20$\,K.

 \citet{2010A&A...512A..54C}  report \Teff\ estimates for GK-type stars using
the IRFM based on BV(RI)$_{\rm C}$JHK$_{\rm S}$ photometry similar to ours.
They argue that the main source of systematic error in the IRFM method is the
conversion of magnitudes to fluxes, i.e. the zero-point of the optical and
infrared magnitude scales. They estimate that a 2\,per~cent error in the
zero-point results in an error of approximately 40\,K in the \Tirfm. Although
our implementation of the IRFM is not the same in detail as that of
\citeauthor{2010A&A...512A..54C}  this estimate of the systematic error
inherent in the method applies equally to our method.

\begin{figure}
\includegraphics[width=0.46\textwidth]{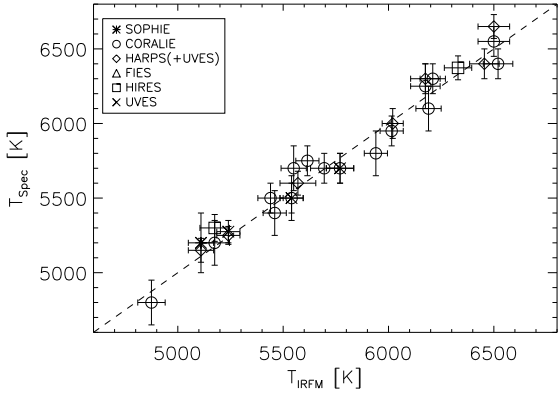}
\caption{Comparision of effective temperature estimates using the IRFM and from spectroscopy.
 \label{teffplot} } 
\end{figure}

 A comparison of the effective temperature estimates \Tirfm\ and \Tspec\ is
shown in Fig.~\ref{teffplot}. The mean value of $\Tspec-\Tirfm$ is
$(-13\pm17)$\,K. It can be seen that the agreement between the two temperature
scales is very good. The $\chi^2$ value for the 1:1 relation shown in
Fig.~\ref{teffplot} is 12.2 for 29 degrees of freedom. This level of agreement
is much better than would be expected given the standard errors quoted for
\Tirfm\ and  \Tspec. However, the uncertainties quoted for \Tspec\ for all the
WASP stars (25 of the 29 \Tspec\ values) include some estimate of the
systematic error in the estimate and these uncertainties are generally quoted
to the nearest 50\,K, e.g., $\pm 100$\,K. If we assume that any systematic error
in  \Tirfm\ is about 40\,K, this suggests that the systematic error in \Tspec\
for WASP stars is likely to be $\la 50$\,K, i.e., similar to the level of
systematic error found by \citeauthor{2010MNRAS.405.1907B} for their \Tspec\
estimates.

 Another quantity sometimes estimated from \Tspec\ is (B$-$V)$_0$, the intrinsic
B$-$V colour. This is used to calculate the chromospheric activity index $\log
R'_{\rm HK}$ \citep{1984ApJ...279..763N}. For WASP stars the calibration of
\cite{2008oasp.book.....G} is used to estimate B$-$V from \Tspec\ (e.g.,
\citealt{2011PASP..123..547M}). For the 19 WASP stars here, we find that this
estimate is accurate to better than 0.03\, magnitudes.  This corresponds to an
additional uncertainty of about 0.05 in the value of $\log R'_{\rm HK}$, which
is small compared to intrinsic decadal variability in this quantity for these
types of stars.

\section{Conclusions}

 We have used \ubvric\ photometry combined with published infrared photometry
to show that the effective temperature estimates for planet host stars
discovered by the WASP survey based an the analysis of the spectrum are
consistent with the infrared flux method effective temperature scale to better
than the quoted standard errors, typically $\pm 100$\,K.

\section*{Acknowledgments}
We thank Francois van Wyk for obtaining a few of the measurements in Table 1.

\bibliographystyle{mn2e}
\bibliography{wasp}

\label{lastpage}

\end{document}